\documentclass{revtex4}  %% REVTeX 4.0
\usepackage[latin1]{inputenc}
\usepackage{amsmath}
\usepackage{amsfonts}
\usepackage{amssymb}
\usepackage{graphicx,color}
\usepackage{epsfig}
\usepackage{epstopdf}
\usepackage{eqnarray}

\begin{document}

%%%%%%%%%%%%%%%%%% title page information %%%%%%%%%%%%%%%%%%
\title{Temporal ghost imaging with twin photons}

\author{S\'{e}verine Denis}
\email[Corresponding author: ]{severine.denis@femto-st.fr}
\affiliation{Institut FEMTO-ST, D\'epartement d'Optique P. M. Duffieux, UMR 6174 CNRS \\ Universit\'e Bourgogne Franche-Comt\'e, 15b Avenue des Montboucons, 25030 Besan\c{c}on - France}
\author{Paul-Antoine Moreau}
\affiliation{Centre for Quantum Photonics, H. H. Wills Physics Laboratory and Department of Electrical and Electronic Engineering, University of Bristol, Merchant Venturers Building, Woodland Road, Bristol BS8 1UB, United Kingdom}
\author{Fabrice Devaux}
\affiliation{Institut FEMTO-ST, D\'epartement d'Optique P. M. Duffieux, UMR 6174 CNRS \\ Universit\'e Bourgogne Franche-Comt\'e, 15b Avenue des Montboucons, 25030 Besan\c{c}on - France}
\author{Eric Lantz}
\affiliation{Institut FEMTO-ST, D\'epartement d'Optique P. M. Duffieux, UMR 6174 CNRS \\ Universit\'e Bourgogne Franche-Comt\'e, 15b Avenue des Montboucons, 25030 Besan\c{c}on - France}

\date{\today}

\begin{abstract}
We use twin photons generated by spontaneous parametric down conversion (SPDC) to perform temporal ghost imaging of a single time signal. The retrieval of a binary signal containing eight bits is performed with an error rate below 1\%. 
\end{abstract}

\maketitle
\section{Introduction}

For the two last decades, ghost imaging has emerged as a kind of magical way to form images of a spatial object, typically a spatially varying transparency, with a Single Point Detector (SPD) that does not have spatial resolution. The initial works used the quantum nature of entanglement of a two-photons state, where photons of a pair are spatially and temporally correlated, to detect temporal coincidences. While one of the photons passing through the object was detected by a photon counter with no spatial resolution, its twin photon was detected with spatial resolution by scanning the transverse plane with a single detector \cite{pittman_optical_1995}, or recently by an intensified charge-coupled device (ICCD) \cite{morris_imaging_2015}. Later, ghost imaging exploiting the temporal correlations of the intensity fluctuations of classical \cite{bennink_two-photon_2002} or pseudothermal light \cite{ferri_high-resolution_2005} was proposed. The ability to retrieve the object with unity contrast seems the only property that belongs to quantum experiments on their own \cite{gatti_correlated_2004}.\\

The extension of ghost imaging to a time object, i.e. a temporally varying transparency, has been recently demonstrated experimentally \cite{ryczkowski_ghost_2016, devaux_computational_2016, devaux_temporal_2016}. In \cite{ryczkowski_ghost_2016}, the light was transmitted through a "time object" and detected with a slow SPD which cannot resolve the time object, while, in the reference arm, the light that did not interact with the temporal object was detected with a fast SPD. Measurements over several thousands copies of the temporal signal were necessary to retrieve a binary signal with a good signal-to-noise ratio. To retrieve a non-reproducible time object using a single shot acquisition, we proposed in \cite{devaux_computational_2016} the exact space-time transposition of computational ghost imaging \cite{shapiro_computational_2008, bromberg_ghost_2009}: a single shot acquisition of the time object was performed by multiplying it with computer-generated random images, ensuring spatial multiplexing of temporal intensity correlations before detecting the sum image with no temporal resolution. While very simple and costless, this method is slow. To increase the speed to a $kHz$ rate, we reported the use of speckle patterns \cite{devaux_temporal_2016} i.e. the temporal transposition of spatial ghost imaging with pseudothermal light \cite{ferri_high-resolution_2005}.\\

In the present paper, we demonstrate temporal ghost imaging with twin photons generated by spontaneous parametric down conversion (SPDC), i.e the temporal transposition of the first ghost imaging experiments \cite{pittman_optical_1995}: while the photons passing through the temporal object are detected by a photon-counting camera with no temporal resolution, their twins do not interact with the object but are detected with temporal and spatial resolution by a second camera. Note that the use of biphotons for temporal imaging has been studied theoretically in \cite{cho_temporal_2012}.

\section{Experimental overview}
\begin{figure}[ht!]
\centering
\includegraphics[width=9.5cm]{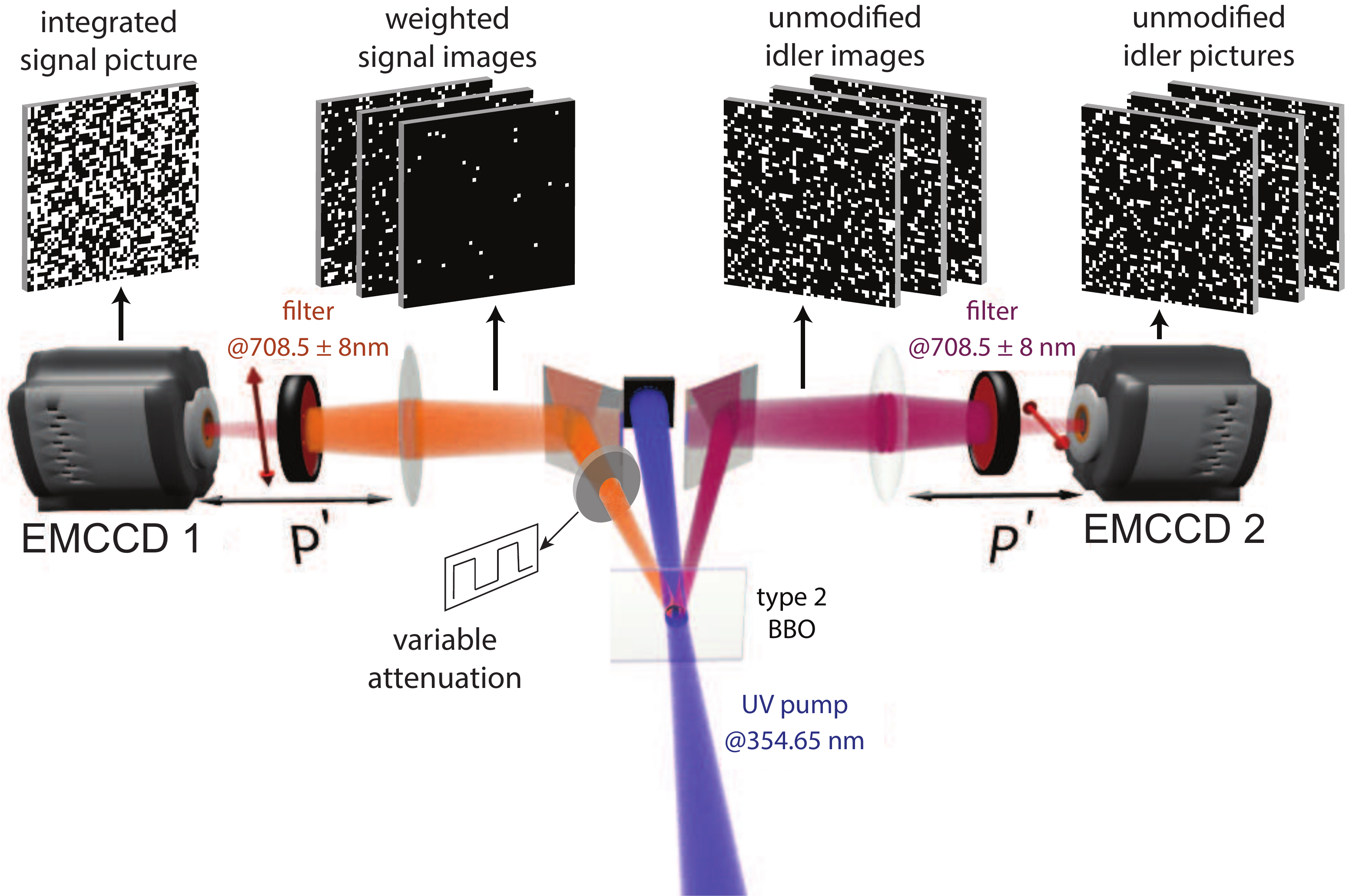}
\caption{Experimental setup used to record the ghost signal and the reference idler images. The red arrows represent the polarisation directions of the signal and idler beams. The variable attenuation, made with a liquid crystal variable retarder and a polariser, is adjusted accordingly.}
\label{fig:setup}
\end{figure}

In the setup represented in Figure \ref{fig:setup}, a type 2 oriented Beta Barium Borate (BBO) nonlinear crystal, with a diameter of 5 mm and a thickness of 0.8 mm, is enlightened over its entire surface with a 354.65 nm UV pulsed laser. From their interaction with the BBO, pump photons are annihilated and generate twin photons that form the signal and idler SPDC beams at 708.5$\pm$8 nm. The filters used have a quasi-rectangular spectrum transmission profile,  centered at 708.5$\pm$8 nm, i.e. not exactly at the degeneracy wavelength of the SPDC.

A variable density made with a liquid crystal variable retarder controlled by a function generator, and followed by a polarizer, is placed on the way of the signal beam. Hence, the signal images are first weighted one by one by the variable density whose values are given by the temporal signal to be retrieved, then summed together, without any temporal resolution, on an electron multiplying charge coupled device  (EMCCD1).  At exactly the same time, the unweighed idler images are acquired one by one by a second camera (EMCCD2) as reference patterns, with for each image $I_n$ a time exposure synchronized with the step $n$ of the time signal.
The signal and idler images are recorded in the image plane of the BBO crystal, in order to ensure a good match between the positions of the signal and idler photons in twin images, even if their wavelengths are slightly different due to the 16 nm width filters. The $512\times512$ pixels EMCCDs (Andor iXon3 897) sensors are cooled at $-100^\circ$C, and ensure a quantum efficiency over 90\% at 708 nm. The photon localizations are recorded by applying a thresholding procedure, (as shown in Figure \ref{fig:setup}). The mean flux is set between $0.10$ and $0.20$ photon per pixel (ph/px) on the sum image, i.e. less on the reference images, in order to minimize the whole number of false detections \cite{lantz_multi-imaging_2008}. \\

 The equivalent quantum efficiency $\eta$ of the setup is given for twin images by the number of detected signal (or idler) photons corresponding to a true pair divided by the total number of detected photons. This parameter takes into account the overall quantum efficiency of the setup, affected by the random absorption of photons by the optical components or no detection by the cameras, but also parasitic fluorescence of the optical components and false decisions during the thresholding procedure. We have shown in \cite{lantz_optimizing_2014} that sources of single photons, like parasitic fluorescence or photons transmitted at the edge of the filters with no transmission at the twin wavelength, have an effect similar to a decrease of the quantum efficiency. Here, the equivalent quantum efficiency of the filters is estimated at 92\% and the combined maximum transmission of the retarder and the polarizer is estimated at $\eta_L=81\%$. Noises from the detector, like readout noise or clock induced charges (CIC) result also in single photoelectrons: either a non genuine photon, due for example to CIC, is detected (false positive error) or a genuine photon is not detected, because the associated level at the output of the multiplication register remains below the threshold (false negative error). This latter case is directly equivalent to a decrease of the quantum efficiency. The former, false positive error, cannot be considered equivalent as a variation of the quantum efficiency since its occurrence does not depend of the light flux \cite{lantz_optimizing_2014}. However, it does result in the creation of single detected photons and   its effect is completely similar to a decrease of the quantum efficiency in integral measurements like that performed in this experiment. The background noises were estimated by recording images with the pump beam off. They are estimated at $0.018~ph/px$ for the signal image and $0.0064~ph/px$ for one of the idler images. 
  \begin{figure}[ht!]
\centering
\includegraphics[width=12cm]{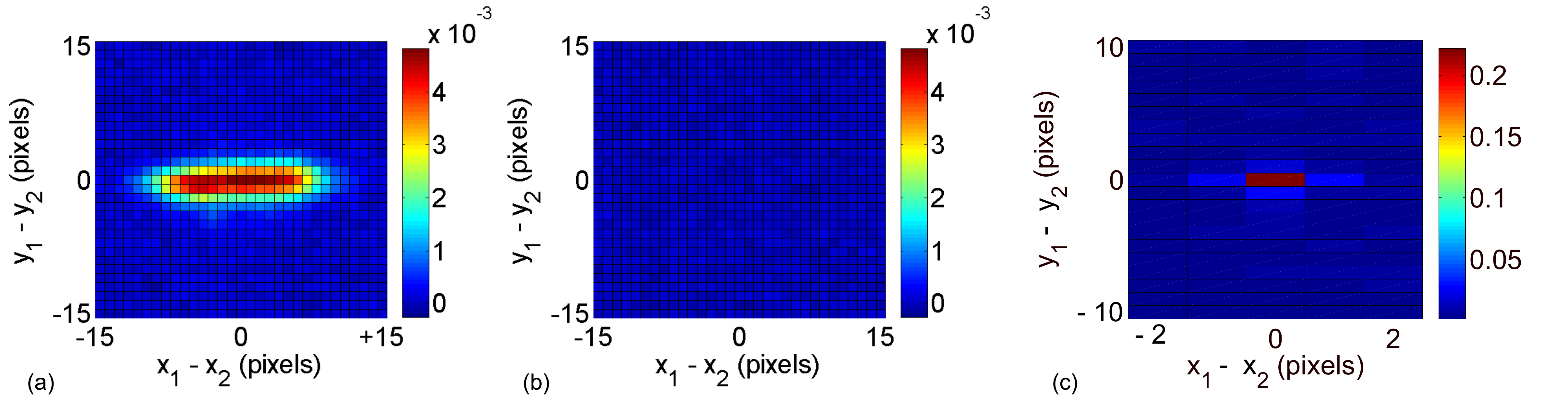}
\caption{$(a)$: mean on 900 realizations of the normalized cross-correlation coefficient of two twin images, without binning.  $(b)$: the same for two independent signal and idler images. $(c)$: the cross-correlation coefficient of figure $(a)$ after a binning of $16\times5$ pixels.}
\label{fig:corr}
\end{figure} 
 The  experimental value of $\eta$ is directly given by the normalized cross-correlation coefficient of the signal and idler pictures. However, in our case, the size of the spatial coherence cells of the SPDC beams does not correspond to the pixels and scales as the inverse of the phase matching angular range \cite{moreau_einstein-podolsky-rosen_2014}. The cross-correlation then displays a peak with a Gaussian like shape that spreads on several pixels. To obtain the full cross-correlation peak on one pixel, a grouping (binning) of B pixels of the cross-correlation figure must be performed. To estimate $\eta$, we recorded one series of 900 twin images, with no time modulation (transmission set to one) and the same integration time on each side. The mean filling ratio $m_s$ was set at $0.04 ph/px$ by adjusting the acquisition time of the cameras. The number of pixels $D=506\times 506$ in each picture used for the cross-correlation is then given by the effective area of the cameras sensors. The average normalized cross-correlation is given in figure \ref{fig:corr}. The size of the spatial coherence cell is estimated at $B=16\times5$ pixels.  However, the binning of the cross-correlation figure is not used here since the average on 900 cross-correlations allows the distinction of the pixels of the peak from those of the cross-correlation background. The integration of the cross-correlation peak gives an overall equivalent quantum efficiency $\eta$ of $30.2\%$.

\section{Retrieval of a ghost time signal}

After acquisition of the weighted and of the reference images, the reconstruction of the time signal is performed by computing the cross-correlation of the signal image and each reference idler image. Here, since we use only one pair of images to determine each cross-correlation coefficient, the cross-correlation peak must be binned in order to be distinguished from the cross-correlation background. The successive cross-correlation coefficients are then plotted over the time to retrieve the shape of the time signal. \\

In an ideal experiment with unity quantum efficiency, a photoelectron detected in the integrated signal image corresponds always to a photon at the same position in one of the reference idler images. However, the random distribution of SPDC photons provides, for different time steps for the signal and the idler or different temporal modes in a single time step, statistically independent photon repartitions. Consequently two signal and idler photons that are not twin can be situated in a coherence cell at the same position. Those photons create accidental matches when cross-correlating the associated signal and idler images. We thus need to consider in the computation of the cross-correlation coefficient a contribution related to accidental coincidences of independent events, but also a fluctuation of the twin and accidental coincidences due to the random nature of the events. The reconstruction of the time signal can be performed properly  in a single operation only if we can distinguish at least two levels in the signal (case of a binary signal). For a Gaussian distribution, this distinction is feasible in 99.3\% of cases if the signal to noise ratio $SNR_n$ associated to the step $n$ verifies :

\begin{equation}
SNR_n = \frac{C_n}{\sigma_n}\geq 2.45~L~T_n
\label{eq:refname1}
\end{equation}

where $\sigma_n$ is the overall standard deviation of the number of coincidences, $L=2$ is the number of levels in the signal, $T_n$ is the binary transmission ($T_n=0$ or $1$) induced by the variable attenuation for the step $n$ of the time signal, and $C_n$ is the mean total number of twin coincidences between the sum picture $S$ and the idler image $I_n$. The value $2.45\times \sigma$ represents the abscissa at which the cumulative density function of a Gaussian distribution takes a value 99.3\%.

The average number of accidental coincidences is shifted to zero by removing in each picture the deterministic shape of the SPDC beams. This step ensures the statistical independence of two non twin images. Two effective methods can be applied here. The first one consists in assimilating the shape of the SPDC beam as a Gaussian, since the intensity of the SPDC beams is proportional to the intensity of the pump beam. Each signal and idler picture is fitted by a Gaussian profile that is then removed from the image. This method is efficient if the experiment is limited to the retrieval of a unique time signal and not repeated \cite{lantz_einstein-podolsky-rosen_2015}. However, if the shape of the beams is not perfectly Gaussian, this method could leave some residual deterministic correlations. The second method consists in recording a large number of images to determine the average deterministic shape of the signal and idler beams, as a calibration of the system. The average shape is then removed from each picture before performing the cross-correlation \cite{moreau_einstein-podolsky-rosen_2014}. If feasible, this last method is slightly more efficient and much more rapid. It  will be  used in the following.  

$C_n$ can be determined from our experimental parameters, after this subtraction, as \cite{lantz_optimizing_2014}:

\begin{equation}
C_n \simeq T_n D (\eta m_i-m_i^2)
\label{eq:refname2}
\end{equation}

where $m_i$ is the mean number of photons (events) per pixel of the idler images. The approximation is valid if the incident photon flux per time step $m_i/\eta$ is much smaller than one, which allows the probability of two photons incident on the same pixel to be neglected.

 $\sigma_n$ takes into account the fluctuations of both twin and accidental coincidences :
 
\begin{equation}
\sigma_n = {(V_{c,n} + V_{a,n})}^{1/2}
\label{eq:refname3}
\end{equation}

where $V_{c,n}$ and $V_{a,n}$ are respectively the variances of the total number of twin and accidental coincidences between $S$ and $I_n$, at the location of the peak. Because of the poissonian distribution of the SPDC pattern, we have directly $V_{c,n} = C_n $. Since $V_{a,n}$ is not related to the number of twin coincidences between the signal integrated image $S$ and one of the idler reference image $I_n$, it does not depend on the step $n$, consequently $V_{a,n}=V_{a}$. Its value can be assessed as follows, in a similar manner as in ref \cite{lantz_optimizing_2014}. We want to assess the fluctuations of $\widehat{cov}(N_s,N_{i_k})$, the estimator of the covariance between one pixel $s$ of the signal picture $S$ and the same pixel $i_k=s$ of the idler picture $I_k$, for independent events (no twin coincidences). $N_s$ and $N_{i_k}$ are the intensities of the $s$ and ${i_k}$ pixels, that are in our case either 1 for one photon  or 0 for no photon. $\widehat{cov}(N_s,N_{i_k})$ is given for each couple $s=i_k$ of the area D by :

\begin{equation}
\begin{split}
\ \widehat{cov}(N_s,N_{i_k}) = \frac{1}{D} \sum \limits_{s=i_k=1}^{D} (N_{s}-\overline{N}_{s}) (N_{i_k}-\overline{N}_{i_k})\\
=\overline{N_{s}N_{i_k}}-\overline{N}_{s}\overline{N}_{i_k}
\end{split}
\label{eq:refname4}
\end{equation}

Because of the independence of the events, $<\widehat{cov}(N_s,N_{i_k})>=0$, where $< >$ stands for the true mean (mathematical expectation). In the second term of Eq.\ref{eq:refname4}, the variance of $\overline{N}_{s}\overline{N}_{i_k}$ is negligible with respect to the variance of $\overline{N_{s}N_{i_k}}$.  The only possible values of $N$ are 0 and 1. Hence we have :
 
\begin{equation}
\begin{split}
\ var(\overline{N_{s}N_{i_k}})=\frac{1}{D}(<(N_{s}N_{i_k})^{2}>-<N_{s}N_{i_k}>^{2})\\
=\frac{1}{D}m_s m_i-(m_s m_i)^{2}\simeq\frac{1}{D}m_s m_i
\end{split}
\label{eq:refname5}
\end{equation}

where $m_s$ and $m_i$ are respectively the true means of the signal and idler images. In the last approximative equality, we assume that $m_s$ and $m_i$ are both $<<1$.\\
 The total number of coincidences is given also by Eq.\ref{eq:refname4}, but without the division by the number of pixels. Hence, we have, if no binning: 

\begin{equation}
\begin{split}
V_{a}=var\left(\sum \limits_{s=i_k=1}^{D} N_{s} N_{i_k}\right)=D~ m_s m_i
\end{split}
\label{eq:refname6}
\end{equation}

 The last step consists in calculating the variance $V_{a}$ of the total number of coincidences between two areas obtained by summing $B$ adjacent pixel values of the correlation image:
 
 \begin{equation}
 \begin{split}
 V_{a}=var\left(\sum \limits_{b=1}^{B}\sum \limits_{s=i_k=1}^{D} N_{s} N_{i_k}\right)=D~ B~ m_s m_i
 \end{split}
 \label{eq:refname7}
 \end{equation}
 
 The signal to noise ratio hence becomes :

 \begin{equation}
SNR_n = \frac{T_n D (\eta m_i-m_i^2)}{(T_n D \eta m_i + D B~ m_s m_i)^{1/2}}
\label{eq:refname10}
\end{equation}

Because of the binning B, the second term of the denominator, due to accidental coincidences, is much greater than the first term, due to the fluctuations of the number of twin coincidences. By neglecting this first term and the second term of the numerator, and by assuming a binary signal with M bits at one, we obtain an approximation of $SNR_n$ as:

\begin{equation}
SNR_n \simeq T_n  \eta\left(\frac{D}{B~ M}\right)^{1/2}
\label{eq:refname11}
\end{equation}

The approximation of Eq.\ref{eq:refname11}, though not very precise (the second term of the numerator is not completely negligible) gives us a practical clue. We have to find an optimal compromise for the binning: increasing $B$ allows the surface of the cross-correlation peak to be entirely covered, resulting in an increase of $\eta$, but at the expense of a decreasing of the number of resolution cells $D/B$ . Experimentally, the lowest error rate has been attained for a binning of 5 pixels on the y axis and 16 pixels on the x axis. This reduced binning unfortunately brings us to ignore the fourth of the coincidences, that are situated on pixels outside the position of the binned peak. The equivalent quantum efficiency thus becomes $\eta=23\%$. \\
By taking into account these values, we have chosen to perform the reconstruction of a time signal of 2 levels (binary) and 8 steps, that should result in a $SNR$ around 6.

\section{Experimental application}
\begin{figure}[ht!]
\centering
\includegraphics[width=16cm]{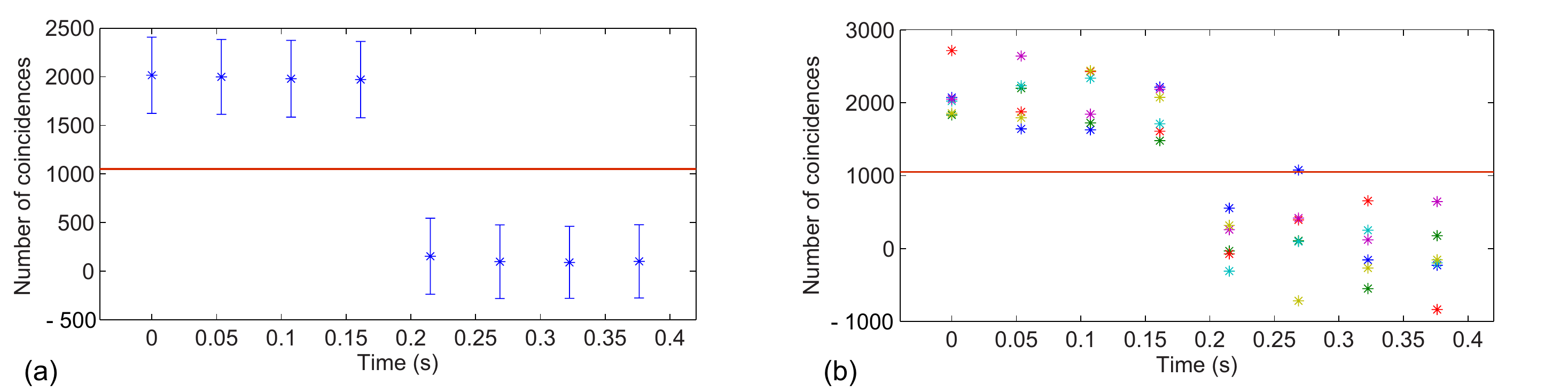}
\caption{(a) Average on 990 reconstructions. The blue dots are the average numbers of coincidences as given in Eq.\ref{eq:refname2}, and the error bars are their standard deviation. (b) Superposition of five reconstructions. The full red line represent the threshold situated at the half average number of coincidences for the "1" steps.}
\label{fig:huit}
\end{figure} 

Although the purpose of the experiment is to benefit of a safe reconstruction on a single operation, the process is repeated 990 times. This allows the estimation of the average experimental $SNR$ on the reconstructed steps. The ghost signal reconstructed here is made of 4 bits at "1" and 4 bits at "0". The reconstruction shown in figure \ref{fig:huit} displays the average cross-correlation coefficient of each step of the time signal, given as a number of coincidences. Let us recall that the average images have been subtracted, meaning that the mean numbers of accidental coincidences have been set to 0, even if these accidental coincidences are the main source of noise, as shown above. The error bars represent the experimental standard deviation of the computed numbers of coincidences associated to each step. 

The $SNR$ of the steps "1" is here equal to $4.9$, while a direct application of Eq.\ref{eq:refname10} gives a $SNR$ of 6.3.  The most important factors that explain this difference between  are:\\ 
- the gaussian shape of the beams result in an effective number of pixels which is smaller than the number $D$ of physical pixels.\\
- In the low light level parts of the image, the detector noises are more important than taken into account by the effective quantum efficiency.\\
- The fluctuations of the pixels in the correlation image are not completely independent, probably because of some smearing. Experimentally, for B=1 the standard deviation of the correlation image has a value outside the twin peak equal to $37.3$, in rather good agreement with its theoretical value $D (m_s m_i)^{1/2}=35.8$. On the other hand, its value of $373$ for B=80 is greater than the expected $320$.

The error rate of $0.7\%$ is in full agreement with the experimentally measured $SNR$. It has been obtained by a simple method of thresholding, where a threshold (red full line in figure \ref{fig:huit}) is placed at the middle between the mean $1$ level and the $0$ one.
  
\section{Conclusion}

We showed in this experiment of temporal ghost imaging that it is possible to reconstruct single sequences of time signals, using quantum correlated photons of SPDC. The number of coherence cells contained in the images determines the available whole length and the error rate in the reconstruction process. The relatively low value of this number, around 3000 here, can be increased in two ways :\\ 
- a thinner crystal allows the decrease of the size of a coherence cell in the image plane by increasing the phase matching range in the Fourier plane,\\
- a wider crystal allows increasing the number of coherence cells in a transverse section. 

However, the conservation of the SPDC gain would require a more powerfull pump beam, as the surface illuminated is larger and the interaction time between the pump pulses and the crystal is smaller. Likewise, more efficient detectors and a reduction of the parasitic fluorescence could result in an increase of the equivalent quantum efficiency. The phase matching constraints explain that performances remain below that obtained with classical means \cite{devaux_computational_2016, devaux_temporal_2016}. Nevertheless, this experiment shows that temporal ghost imaging can be performed by using either twin photons or classical correlations, just as for spatial ghost imaging.
  
\section*{Funding}
This work was supported by the Labex ACTION program (ANR-11-LABX-0001-01).

%\bibliography{GhostImaging}

\begin{thebibliography}{15}
\expandafter\ifx\csname natexlab\endcsname\relax\def\natexlab#1{#1}\fi
\expandafter\ifx\csname bibnamefont\endcsname\relax
  \def\bibnamefont#1{#1}\fi
\expandafter\ifx\csname bibfnamefont\endcsname\relax
  \def\bibfnamefont#1{#1}\fi
\expandafter\ifx\csname citenamefont\endcsname\relax
  \def\citenamefont#1{#1}\fi
\expandafter\ifx\csname url\endcsname\relax
  \def\url#1{\texttt{#1}}\fi
\expandafter\ifx\csname urlprefix\endcsname\relax\def\urlprefix{URL }\fi
\providecommand{\bibinfo}[2]{#2}
\providecommand{\eprint}[2][]{\url{#2}}

\bibitem[{\citenamefont{Pittman et~al.}(1995)\citenamefont{Pittman, Shih,
  Strekalov, and Sergienko}}]{pittman_optical_1995}
\bibinfo{author}{\bibfnamefont{T.~B.} \bibnamefont{Pittman}},
  \bibinfo{author}{\bibfnamefont{Y.~H.} \bibnamefont{Shih}},
  \bibinfo{author}{\bibfnamefont{D.~V.} \bibnamefont{Strekalov}},
  \bibnamefont{and} \bibinfo{author}{\bibfnamefont{A.~V.}
  \bibnamefont{Sergienko}}, "Optical imaging by means of two-photon quantum entanglement",\bibinfo{journal}{Phys. Rev. A}
  \textbf{\bibinfo{volume}{52}}, \bibinfo{pages}{R3429} (\bibinfo{year}{1995}),
  \urlprefix\url{http://link.aps.org/doi/10.1103/PhysRevA.52.R3429}.

\bibitem[{\citenamefont{Morris et~al.}(2015)\citenamefont{Morris, Aspden, Bell,
  Boyd, and Padgett}}]{morris_imaging_2015}
\bibinfo{author}{\bibfnamefont{P.~A.} \bibnamefont{Morris}},
  \bibinfo{author}{\bibfnamefont{R.~S.} \bibnamefont{Aspden}},
  \bibinfo{author}{\bibfnamefont{J.~E.~C.} \bibnamefont{Bell}},
  \bibinfo{author}{\bibfnamefont{R.~W.} \bibnamefont{Boyd}}, \bibnamefont{and}
  \bibinfo{author}{\bibfnamefont{M.~J.} \bibnamefont{Padgett}}, "Imaging with a small number of photons",
  \bibinfo{journal}{Nat Commun} \textbf{\bibinfo{volume}{6}},
  \bibinfo{pages}{5913} (\bibinfo{year}{2015}),
  \urlprefix\url{http://www.nature.com/ncomms/2015/150105/ncomms6913/full/ncomms6913.html}.

\bibitem[{\citenamefont{Bennink et~al.}(2002)\citenamefont{Bennink, Bentley,
  and Boyd}}]{bennink_two-photon_2002}
\bibinfo{author}{\bibfnamefont{R.~S.} \bibnamefont{Bennink}},
  \bibinfo{author}{\bibfnamefont{S.~J.} \bibnamefont{Bentley}},
  \bibnamefont{and} \bibinfo{author}{\bibfnamefont{R.~W.} \bibnamefont{Boyd}}, "Two-Photon Coincidence Imaging with a Classical Source",
  \bibinfo{journal}{Phys. Rev. Lett.} \textbf{\bibinfo{volume}{89}},
  \bibinfo{pages}{113601} (\bibinfo{year}{2002}),
  \urlprefix\url{http://link.aps.org/doi/10.1103/PhysRevLett.89.113601}.

\bibitem[{\citenamefont{Ferri et~al.}(2005)\citenamefont{Ferri, Magatti, Gatti,
  Bache, Brambilla, and Lugiato}}]{ferri_high-resolution_2005}
\bibinfo{author}{\bibfnamefont{F.}~\bibnamefont{Ferri}},
  \bibinfo{author}{\bibfnamefont{D.}~\bibnamefont{Magatti}},
  \bibinfo{author}{\bibfnamefont{A.}~\bibnamefont{Gatti}},
  \bibinfo{author}{\bibfnamefont{M.}~\bibnamefont{Bache}},
  \bibinfo{author}{\bibfnamefont{E.}~\bibnamefont{Brambilla}},
  \bibnamefont{and} \bibinfo{author}{\bibfnamefont{L.~A.}
  \bibnamefont{Lugiato}}, "High-Resolution Ghost Image and Ghost Diffraction Experiments with Thermal Light", \bibinfo{journal}{Phys. Rev. Lett.}
  \textbf{\bibinfo{volume}{94}}, \bibinfo{pages}{183602}
  (\bibinfo{year}{2005}),
  \urlprefix\url{http://link.aps.org/doi/10.1103/PhysRevLett.94.183602}.

\bibitem[{\citenamefont{Gatti et~al.}(2004)\citenamefont{Gatti, Brambilla,
  Bache, and Lugiato}}]{gatti_correlated_2004}
\bibinfo{author}{\bibfnamefont{A.}~\bibnamefont{Gatti}},
  \bibinfo{author}{\bibfnamefont{E.}~\bibnamefont{Brambilla}},
  \bibinfo{author}{\bibfnamefont{M.}~\bibnamefont{Bache}}, \bibnamefont{and}
  \bibinfo{author}{\bibfnamefont{L.~A.} \bibnamefont{Lugiato}}, "Correlated imaging, quantum and classical",
  \bibinfo{journal}{Physical Review A} \textbf{\bibinfo{volume}{70}},
  \bibinfo{pages}{013802} (\bibinfo{year}{2004}),
  \urlprefix\url{http://link.aps.org/doi/10.1103/PhysRevA.70.013802}.

\bibitem[{\citenamefont{Ryczkowski et~al.}(2016)\citenamefont{Ryczkowski,
  Barbier, Friberg, Dudley, and Genty}}]{ryczkowski_ghost_2016}
\bibinfo{author}{\bibfnamefont{P.}~\bibnamefont{Ryczkowski}},
  \bibinfo{author}{\bibfnamefont{M.}~\bibnamefont{Barbier}},
  \bibinfo{author}{\bibfnamefont{A.~T.} \bibnamefont{Friberg}},
  \bibinfo{author}{\bibfnamefont{J.~M.} \bibnamefont{Dudley}},
  \bibnamefont{and} \bibinfo{author}{\bibfnamefont{G.}~\bibnamefont{Genty}}, "Ghost imaging in the time domain",
  \bibinfo{journal}{Nat Photon} \textbf{\bibinfo{volume}{10}},
  \bibinfo{pages}{167} (\bibinfo{year}{2016}), ISSN \bibinfo{issn}{1749-4885},
  \urlprefix\url{http://www.nature.com/nphoton/journal/v10/n3/full/nphoton.2015.274.html}.

\bibitem[{\citenamefont{Devaux et~al.}(2016{\natexlab{a}})\citenamefont{Devaux,
  Moreau, Denis, and Lantz}}]{devaux_computational_2016}
\bibinfo{author}{\bibfnamefont{F.}~\bibnamefont{Devaux}},
  \bibinfo{author}{\bibfnamefont{P.-A.} \bibnamefont{Moreau}},
  \bibinfo{author}{\bibfnamefont{S.}~\bibnamefont{Denis}}, \bibnamefont{and}
  \bibinfo{author}{\bibfnamefont{E.}~\bibnamefont{Lantz}}, "Computational temporal ghost imaging",
  \bibinfo{journal}{Optica} \textbf{\bibinfo{volume}{3}}, \bibinfo{pages}{698}
  (\bibinfo{year}{2016}{\natexlab{a}}), ISSN \bibinfo{issn}{2334-2536},
  \urlprefix\url{http://www.osapublishing.org/abstract.cfm?uri=optica-3-7-698}.

\bibitem[{\citenamefont{Devaux et~al.}(2016{\natexlab{b}})\citenamefont{Devaux,
  Huy, Moreau, Denis, and Lantz}}]{devaux_temporal_2016}
\bibinfo{author}{\bibfnamefont{F.}~\bibnamefont{Devaux}},
  \bibinfo{author}{\bibfnamefont{K.~P.} \bibnamefont{Huy}},
  \bibinfo{author}{\bibfnamefont{P.-A.} \bibnamefont{Moreau}},
  \bibinfo{author}{\bibfnamefont{S.}~\bibnamefont{Denis}}, \bibnamefont{and}
  \bibinfo{author}{\bibfnamefont{E.}~\bibnamefont{Lantz}}, "Temporal ghost imaging with pseudo-thermal speckle light",
  \bibinfo{journal}{arXiv:1609.05465 [physics]}
  (\bibinfo{year}{2016}{\natexlab{b}}), \bibinfo{note}{arXiv: 1609.05465},
  \urlprefix\url{http://arxiv.org/abs/1609.05465}.

\bibitem[{\citenamefont{Shapiro}(2008)}]{shapiro_computational_2008}
\bibinfo{author}{\bibfnamefont{J.~H.} \bibnamefont{Shapiro}}, "Computational ghost imaging",
  \bibinfo{journal}{Phys. Rev. A} \textbf{\bibinfo{volume}{78}},
  \bibinfo{pages}{061802} (\bibinfo{year}{2008}),
  \urlprefix\url{http://link.aps.org/doi/10.1103/PhysRevA.78.061802}.

\bibitem[{\citenamefont{Bromberg et~al.}(2009)\citenamefont{Bromberg, Katz, and
  Silberberg}}]{bromberg_ghost_2009}
\bibinfo{author}{\bibfnamefont{Y.}~\bibnamefont{Bromberg}},
  \bibinfo{author}{\bibfnamefont{O.}~\bibnamefont{Katz}}, \bibnamefont{and}
  \bibinfo{author}{\bibfnamefont{Y.}~\bibnamefont{Silberberg}}, "Ghost imaging with a single detector",
  \bibinfo{journal}{Phys. Rev. A} \textbf{\bibinfo{volume}{79}},
  \bibinfo{pages}{053840} (\bibinfo{year}{2009}),
  \urlprefix\url{http://link.aps.org/doi/10.1103/PhysRevA.79.053840}.

\bibitem[{\citenamefont{Cho and Noh}(2012)}]{cho_temporal_2012}
\bibinfo{author}{\bibfnamefont{K.}~\bibnamefont{Cho}} \bibnamefont{and}
  \bibinfo{author}{\bibfnamefont{J.}~\bibnamefont{Noh}}, "Temporal ghost imaging of a time object, dispersion cancelation, and nonlocal time lens with bi-photon state",
  \bibinfo{journal}{Optics Communications} \textbf{\bibinfo{volume}{285}},
  \bibinfo{pages}{1275} (\bibinfo{year}{2012}), ISSN \bibinfo{issn}{0030-4018},
  \urlprefix\url{http://www.sciencedirect.com/science/article/pii/S0030401811011242}.

\bibitem[{\citenamefont{Lantz et~al.}(2008)\citenamefont{Lantz, Blanchet,
  Furfaro, and Devaux}}]{lantz_multi-imaging_2008}
\bibinfo{author}{\bibfnamefont{E.}~\bibnamefont{Lantz}},
  \bibinfo{author}{\bibfnamefont{J.-L.} \bibnamefont{Blanchet}},
  \bibinfo{author}{\bibfnamefont{L.}~\bibnamefont{Furfaro}}, \bibnamefont{and}
  \bibinfo{author}{\bibfnamefont{F.}~\bibnamefont{Devaux}}, "Multi-imaging and Bayesian estimation for photon counting with EMCCDs",
  \bibinfo{journal}{Monthly Notices of the Royal Astronomical Society}
  \textbf{\bibinfo{volume}{386}}, \bibinfo{pages}{2262} (\bibinfo{year}{2008}),
  ISSN \bibinfo{issn}{0035-8711, 1365-2966},
  \urlprefix\url{http://mnras.oxfordjournals.org/content/386/4/2262}.
  
  \bibitem[{\citenamefont{Lantz et~al.}(2014)\citenamefont{Lantz, Moreau, and
  Devaux}}]{lantz_optimizing_2014}
\bibinfo{author}{\bibfnamefont{E.}~\bibnamefont{Lantz}},
  \bibinfo{author}{\bibfnamefont{P.-A.} \bibnamefont{Moreau}},
  \bibnamefont{and} \bibinfo{author}{\bibfnamefont{F.}~\bibnamefont{Devaux}}, "Optimizing the signal-to-noise ratio in the measurement of photon pairs with detector arrays",
  \bibinfo{journal}{Physical Review A} \textbf{\bibinfo{volume}{90}},
  \bibinfo{pages}{063811} (\bibinfo{year}{2014}),
  \urlprefix\url{http://link.aps.org/doi/10.1103/PhysRevA.90.063811}.

\bibitem[{\citenamefont{Moreau et~al.}(2014)\citenamefont{Moreau, Devaux, and
  Lantz}}]{moreau_einstein-podolsky-rosen_2014}
\bibinfo{author}{\bibfnamefont{P.-A.} \bibnamefont{Moreau}},
  \bibinfo{author}{\bibfnamefont{F.}~\bibnamefont{Devaux}}, \bibnamefont{and}
  \bibinfo{author}{\bibfnamefont{E.}~\bibnamefont{Lantz}}, "Einstein-Podolsky-Rosen Paradox in Twin Images"
  \bibinfo{journal}{Physical Review Letters} \textbf{\bibinfo{volume}{113}},
  \bibinfo{pages}{160401} (\bibinfo{year}{2014}),
  \urlprefix\url{http://link.aps.org/doi/10.1103/PhysRevLett.113.160401}.

\bibitem[{\citenamefont{Lantz et~al.}(2015)\citenamefont{Lantz, Denis, Moreau,
  and Devaux}}]{lantz_einstein-podolsky-rosen_2015}
\bibinfo{author}{\bibfnamefont{E.}~\bibnamefont{Lantz}},
  \bibinfo{author}{\bibfnamefont{S.}~\bibnamefont{Denis}},
  \bibinfo{author}{\bibfnamefont{P.-A.} \bibnamefont{Moreau}},
  \bibnamefont{and} \bibinfo{author}{\bibfnamefont{F.}~\bibnamefont{Devaux}}, "Einstein-Podolsky-Rosen paradox in single pairs of images",
  \bibinfo{journal}{Optics Express} \textbf{\bibinfo{volume}{23}},
  \bibinfo{pages}{26472} (\bibinfo{year}{2015}), ISSN
  \bibinfo{issn}{1094-4087},
  \urlprefix\url{http://www.osapublishing.org/abstract.cfm?uri=oe-23-20-26472}.

\end{thebibliography}

%\title{"Optical imaging by means of two-photon quantum entanglement"},

\end{document}